# Bohr Hamiltonian for $\gamma = 0°$ with Davidson Potential

*I. YIGITOGLU and M. GOKBULUT*

*Department of Physics, Faculty of Arts and Sciences, Gaziosmanpaşa University,*

*60250, Tokat, Turkey*

*ibrahim.yigitoglu@gop.edu.tr, melek.kgb@gmail.com*

A $\gamma$-rigid solution of the Bohr Hamiltonian is derived for $\gamma = 0°$ utilizing the Davidson potential in the $\beta$ variable. This solution is going to be called X(3)-D. The energy eigenvalues and wave functions are obtained by using an analytic method which has been developed by Nikiforov and Uvarov. BE(2) transition rates are calculated. A variational procedure is applied to energy ratios to determine whether or not the X(3) model is located at the critical point between spherical and deformed nuclei.

## I. Introduction

The studies describing analytically the critical point at the shape-phase transitions between different dynamical symmetries and enlightening structural properties in atomic nuclei with experimental evidence have been started with the introduction of two new critical point symmetries, called E(5) [1] (between U(5) vibrational and O(6) $\gamma$-unstable nuclei) and X(5) [2] (between U(5) vibrational and SU(3) axially deformed nuclei). The E(5) symmetry is a $\gamma$ independent exact solution of the Bohr Hamiltonian [3], while the X(5) symmetry is an approximate solution for $\gamma \approx 0°$.

The method is based on constructing the Bohr Hamiltonian, choosing different types of potentials such as Morse [4], Kratzer [5-7], Coulomb [5,6], Davidson [5,8], Eckart [9], Manning-Rosen [10], Killingbeck [11] and solving the eigenfunction-eigenvalue problem in search for the quadropole collective dynamics of nuclei. The geometric potentials admitting analytical solutions for the Schrödinger equation belong into two groups. The potentials in the first group depend on both $\beta$ and $\gamma$ and can be written in the form $V(\beta,\gamma) = V(\beta) + V(\gamma)$ [2,12], where the separation of variables can be done approximately, while the potentials in the second group can be written in the form $V(\beta,\gamma) = V(\beta) + V(\gamma)/\beta^2$ [13,14], where the separation of variables is done exactly.

Microscopic studies [15-17] indicate that the potential at the shape phase transition point between two dynamical symmetries in atomic nuclei should be flat. Therefore an infinite square-well potential with respect to the $\beta$ variable is used in the E(5) and X(5) symmetries.

The sequence of potentials studied in [18,19] give the opportunity to approach the E(5) and X(5) symmetries starting from U(5). Davidson type potentials [20], having a minimum at $\beta \neq 0$, are good candidates for approaching the E(5) and X(5) symmetries starting respectively from the O(6) and SU(3) limiting structures. This gives rise to exact solutions which cover all the way from U(5) to O(6) and from U(5) to SU(3). Moreover, using a variational procedure with Davidson potential the physical quantities at the critical point can be obtained [21-23].

The starting idea behind this study is to obtain a $\gamma$-rigid version of the X(5) model constructed by assuming the nucleus to be $\gamma$-rigid, as in the Davydov and Chaban approach [24], and by fixing $\gamma = 0°$, corresponding to the axially symmetric rotor case. The model obtained in this way is called X(3) [25]. Recently three new studies have been performed in this $\gamma$-rigid regime. In two of these [26,27], an infinite square well potential is used in the $\beta$ part of the Schrödinger equation within the minimal length formalism. In the third one [28], two $\gamma$-rigid solutions are obtained for $\gamma = 0°$ and for $\gamma = 30°$ using a sextic potential in the $\beta$ part.

In the present study a version of the X(3) model is introduced by using the Davidson potential [20] in the $\beta$-part of the Schrödinger equation. This solution is going to be called the X(3)-D model. The energy eigenvalues and wave functions are obtained by using an analytical method which has been developed by Nikiforov and Uvarov [29]. BE(2) transitions rates are calculated. A variational procedure is applied for recovering the ground state band energies of the X(3) model predictions, in order to determine whether or not the X(3) model is located at the critical point.

**II. The X(3)-D solution**

The original Bohr Hamiltonian lives in a five dimensional space,

$$H = -\frac{\hbar^2}{2B}\left[\frac{1}{\beta^4}\frac{\partial}{\partial \beta}\beta^4\frac{\partial}{\partial \beta} + \frac{1}{\beta^2}\frac{1}{\sin 3\gamma}\frac{\partial}{\partial \gamma}\sin 3\gamma\frac{\partial}{\partial \gamma} - \frac{1}{4\beta^2}\sum_\kappa \frac{Q_\kappa^2}{\sin^2\left(\gamma - \frac{2\pi}{3}\kappa\right)}\right] + V(\beta,\gamma). \quad (1)$$

Here B is the mass parameter, $\beta$ and $\gamma$ are the collective coordinates and $Q_\kappa(\kappa=1,2,3)$ are the components of angular momentum in the intrinsic frame.

Considering the axially symmetric prolate case for $\gamma = 0°$ [25], it is clearly seen that the motion is characterized by three collective variables $(\beta,\theta,\phi)$ and the Bohr Hamiltonian takes the form

$$H = -\frac{\hbar^2}{2B}\left[\frac{1}{\beta^2}\frac{\partial}{\partial \beta}\beta^2\frac{\partial}{\partial \beta} + \frac{1}{3\beta^2}\frac{1}{\sin\theta}\frac{\partial}{\partial \theta}\sin\theta\frac{\partial}{\partial \theta} + \frac{1}{3\beta^2}\frac{1}{\sin^2\theta}\frac{\partial^2}{\partial \phi^2}\right] + V(\beta). \quad (2)$$

The wave function is

$$\psi(\beta,\theta,\varphi) = F(\beta)Y_{LM}(\theta,\phi). \quad (3)$$

Here $Y_{LM}(\theta,\phi)$ are the spherical harmonics.

The Schrödinger equation can be separated into $\beta$ and angular parts.

$$\left[\frac{1}{\beta^2}\frac{d}{d\beta}\beta^2\frac{d}{d\beta} + \left(\varepsilon - u(\beta) - \frac{L(L+1)}{3\beta^2}\right)\right]F(\beta) = 0, \quad (4)$$

$$-\left[\frac{1}{\sin\theta}\frac{\partial}{\partial\theta}\sin\theta\frac{\partial}{\partial\theta}+\frac{1}{\sin^2\theta}\frac{\partial^2}{\partial\phi^2}\right]Y_{LM}(\theta,\phi)=L(L+1)Y_{LM}(\theta,\phi).\tag{5}$$

Here $L$ is the angular momentum quantum number and reduced energies $\varepsilon=2BE/\hbar^2$ and reduced potentials $u=2BV/\hbar^2$ have been used.

In the X(3) model, the $\beta$ part is solved by taking an infinite square well potential, as it is done in the X(5) model. In the present work, the $\beta$ part is solved for the Davidson potential. Inserting the Davidson potential [20]

$$u(\beta)=\beta^2+\frac{\beta_o^4}{\beta^2}\tag{6}$$

into Eq. (4), the "radial" equation can be rewritten as

$$\frac{d^2F(\beta)}{d\beta^2}+\frac{2}{\beta}\frac{dF(\beta)}{d\beta}+\left[\varepsilon-\frac{L(L+1)}{3\beta^2}-\beta^2-\frac{\beta_o^4}{\beta^2}\right]F(\beta)=0.\tag{7}$$

In order to solve Eq. (7), one needs to transform it into the Nikiforov-Uvarov (NU) equation form. The NU equation reads

$$\psi''(z)+\frac{\tilde{\tau}(z)}{\sigma(z)}\psi'(z)+\frac{\tilde{\sigma}(z)}{\sigma^2(z)}\psi(z)=0,\tag{8}$$

where $\sigma(z)$ and $\tilde{\sigma}(z)$ are at most second degree polynomials, while $\tilde{\tau}(z)$ is a first degree polynomial. For this purpose the $\beta^2=z$ change of variable is applied to Eq. (7) and

$$\frac{d^2F(z)}{dz^2}+\frac{3}{2z}\frac{dF(z)}{dz}+\frac{1}{4z^2}\left[-z^2+\varepsilon z-\alpha\right]F(z)=0\tag{9}$$

is obtained. We compared Eqs. (8) and (9) and then we determined the parametric polynomials as follows

$$\tilde{\tau}(z) = 3, \qquad \sigma(z) = 2z, \qquad \tilde{\sigma}(z) = -z^2 + \varepsilon z - \alpha \qquad (10)$$

where $\alpha = \beta_o^4 + \dfrac{L(L+1)}{3}$. The functions and the parameters required for this method are defined as follows

$$\begin{aligned}
&\pi(z) = \sigma'(z) - \tilde{\tau}(z)/2 \pm \left[ \left( \sigma'(z) - \tilde{\tau}(z)/2 \right)^2 - \tilde{\sigma}(z) + k\sigma(z) \right]^{1/2}, \\
&\tau(z) = \tilde{\tau}(z) + 2\pi(z), \\
&\lambda = k + \pi'(z), \\
&\lambda_n = -n\tau' - \dfrac{n(n-1)}{2} \sigma'' \quad n = 0,1,2,...
\end{aligned} \qquad (11)$$

For $\lambda = \lambda_n$, one obtains the energy eigenvalues. By using the parametric polynomials, we get the equalities

$$\begin{aligned}
&k_- = \varepsilon/2 - \sqrt{\alpha + 1/4}, \\
&\pi(z) = -1/2 - \left[ z - \sqrt{\alpha + 1/4} \right], \\
&\lambda = \varepsilon/2 - \sqrt{\alpha + 1/4} - 1, \\
&\lambda_n = 2n.
\end{aligned} \qquad (12)$$

Finally, we obtained the energy equation

$$E_{n,L} = 2n + 1 + \sqrt{\beta_o^4 + \dfrac{L(L+1)}{3} + \dfrac{1}{4}}, \qquad (13)$$

where $n$ is the usual oscillator quantum number, $L$ is the angular momentum quantum number, and $\beta_o$ corresponds to the position of the minimum of the potential. In Eq. (13), $n=0$ corresponds to the ground state band (gsb).

The eigenfunction is written in the form of $F(z) \to \psi(z) = \varphi(z) y_n(z)$. Following the NU method

$$\frac{\varphi'(z)}{\varphi(z)} = \frac{\pi(z)}{\sigma(z)},$$

$$y_n = \frac{B_n}{\rho(z)} \frac{d^n}{dz^n} \left[\sigma^n(z)\rho(z)\right], \qquad (14)$$

$$\left[\sigma(z)\rho(z)\right]' = \tau(z)\rho(z).$$

We have calculated $\varphi(z)$ and $y_n(z)$ and the eigenfunction is written with respect to the $\beta$ variable as

$$F(\beta) = C_N \beta^a e^{-\beta^2/2} L_n^{a+1/2}(\beta^2), \qquad (15)$$

where $C_N$ is the normalization constant, $a = -\frac{1}{2} + \sqrt{\beta_o^4 + \frac{L(L+1)}{3} + \frac{1}{4}}$, and $L_n$ are the Laguerre polynomials.

The $C_N$ normalization constant is obtained from the condition $\int_0^\infty F^2(\beta)\beta^2 d\beta = 1$. Then the final expression for the eigenfunction is

$$F(\beta) = \left[\frac{2}{\Gamma(a+\frac{3}{2})\binom{n+a+1/2}{n}}\right]^{1/2} \beta^a e^{-\beta^2/2} L_n^{a+1/2}(\beta^2). \qquad (16)$$

The $R_L$ values play an important role in investigating the structural evolution. They are defined as

$$R_L = \frac{E_{0,L} - E_{0,0}}{E_{0,2} - E_{0,0}}. \tag{17}$$

## III. BE2 Transition Rates

The general form of the quadropole operator is

$$T_\mu^{(E2)} = t\beta \left[ D_{\mu,0}^{2*}(\Omega)\cos\gamma + \frac{1}{\sqrt{2}}\left[ D_{\mu,2}^{2*}(\Omega) + D_{\mu,-2}^{2*}(\Omega)\right]\sin\gamma \right], \tag{18}$$

where $t$ denotes a scalar factor and $\Omega$ the Euler angles.

The quadropole operator for $\gamma = 0$ is

$$T_\mu^{(E2)}(\beta) = t\beta \sqrt{\frac{4\pi}{5}} Y_{2\mu}(\theta,\phi). \tag{19}$$

Then the $B(E2)$ rates are

$$B(E2; sL \to s'L') = t^2 \left(C_{L0,20}^{L'0}\right)^2 I_{sL;s'L'}^2, \tag{20}$$

where $C_{L0,20}^{L'0}$ are Clebsch-Gordan coeffients and the integrals are

$$I_{sL \to s'L'} = \int_0^{\beta_W} \beta F_{sL}(\beta) F_{s'L'}(\beta) \beta^2 d\beta. \tag{21}$$

## IV. Numerical Results

The lowest bands for the X(3)-D model can be seen for various values of the parameter $\beta_o$ in Table I. The ground state band is characterized by $(n=0, s=1)$. The $\beta_1$ band is characterized by $(n=1, s=2)$, while the $\beta_2$ band is characterized by $(n=2, s=3)$. The X(3) and X(3)-D spectra can be compared through the correspondence $n = s-1$. The energy levels are represented by $L_{s,n}$. The energy levels of all bands are normalized to the energy of the lowest excited $2_{1,0}$ level. In addition, the energy levels resulting from the variational procedure are reported (labelled by "var"), along with the parameter values $\beta_{o,m}$ at which they are obtained.

The ground state band energies are shown in Figure 1. It is clear that the energy levels for $\beta_o = 0$ approach the U(5) vibrational limit, while $\beta_o \to \infty$ corresponds to the SU(3) rotational limit. The spectra obtained for $\beta_o = 1.5$ and $\beta_o = 2$ are similar to the X(3) and X(5) model predictions respectively. It can be seen that with increasing $\beta_o$ values the way from U(5) to SU(3) is spanned.

**Figure 1**. The energy levels vs angular momentum $L$ are shown for different $\beta_o$ values and are compared with the U(5), SU(3), X(5) and X(3) predictions. All energy levels are normalized to the lowest excited state.

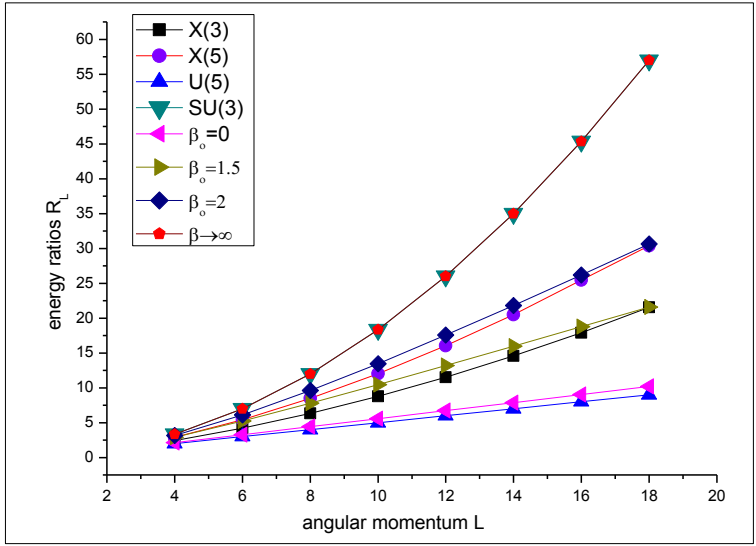

The variational procedure can be used as a tool for determining the behavior of the physical quantities at the point of shape-phase transitions in nuclei. The method is based on determining the parameter value that corresponds to the maximum increase of a physical quantity at the critical point, while using a potential with one parameter spanning the region between the two limiting structures.

The $R_L$ ratios are structural signatures that serve for searching the shape-phase transition and the critical point between two limiting symmetries. The determination of the $\beta_o$ values at which the rate of change of $R_L$ becomes maximum is crucial, since the structural evolution at the critical point changes rapidly. Therefore we applied the variational procedure and compared the obtained results to the X(3) model predictions to determine whether or not the X(3) model is located at the critical point between the spherical vibrator U(5) and the axially symmetric prolate rotor SU(3).

The $R_L$ curves exhibited in Figure 2 show the evolution of nuclear structure from the U(5) symmetry on the left hand side to the SU(3) symmetry on the right hand side. The $R_L$ ratios increase with the $\beta_o$ values. The curve has the steepest increase at the point $\beta_{o,m}$, where the first derivative acquires its maximum value, while the second derivative $d^2 R_L / d\beta_o^2$ vanishes at this point.

**Figure 2**. The $R_L$ ratios for $L = 4, 12, 18$ and their derivatives $dR_L / d\beta_o$ versus the $\beta_o$ parameter values obtained using the Davidson potential in the X(3)-D model.

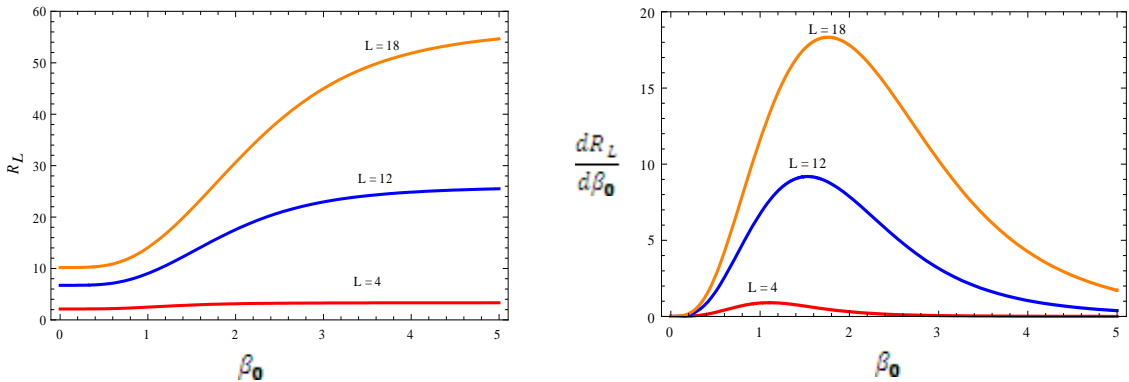

Numerical results for $\beta_{o,m}$ and the corresponding $R_L$ ratios (labelled as var) obtained through the variational procedure are shown in Table I. We see that the output of the variational procedure (the "var" column) is quite similar to the X(3) results, although the agreement is not as strong as in the studies where the same comparison is made between the variational procedure and the E(5) and X(5) model predictions [22]. This discrepancy may be attributed to the $\gamma$-rigid nature of the X(3) model, where $\gamma$ is a parameter and not a variable.

The ground state band energies obtained for different $\beta_o$ values are compared with the existing experimental data of the $^{172-180}$Os, $^{104}$Ru, $^{148}$Nd, $^{186}$Pt, $^{196}$Pt, $^{120}$Xe, $^{126}$Xe, $^{154}$Gd and $^{156}$Dy isotopes in Figure 3. The X(5) and X(3) model predictions are also shown in Figure 3 for comparison.

The $^{172}$Os ground state band (gsb) energies are in good agreement with X(3)-D (with $\beta_o$ =1.310) and X(3) model predictions up to $L=12$. The X(3)-D model predictions with increasing $\beta_o$ values move to the X(5) model side and become consistent with the X(5) model and the $^{176}$Os data for $\beta_o$ =1.863. The $^{178-180}$Os data are also lying close to X(5), described by X(3)-D with $\beta_o$ =1.910 and 2.0 respectively.

The X(3) model gsb is in good agreement with $^{104}$Ru up to $L=10$ and with $^{148}$Nd up to $L=8$. Both the $^{104}$Ru and $^{148}$Nd isotopes show better agreement up to higher L with X(3)-D (with $\beta_o$ =1.210 and 1.115 respectively).

The same behavior can be observed for the Xe isotopes. The X(3) model gsb is in good agreement with $^{120}$Xe up to $L=12$ and with $^{126}$Xe up to L=10, while better agreement up to higher L can be seen for X(3)-D (with $\beta_o$ =1.300 and 1.204 respectively).

Similar behavior can be observed for the Pt isotopes. Good agreement with $^{186}$Pt and $^{196}$Pt is seen for X(3)-D with $\beta_o$ =1.400 and 1.301 respectively.

The well known X(5) model candidates $^{154}$Gd and $^{156}$Dy are in good agreement with the X(3)-D model for $\beta_o$ =1.980 and $\beta_o$ =1.905 respectively. This is a piece of evidence on how the X(3)-D model spans the path from the U(5) symmetry to the SU(3) symmetry, as it is also seen in Figure 2.

**Figure 3**. The X(3)-D model ground state band energy predictions obtained for different $\beta_o$ values are compared with the data for $^{172-180}$Os, $^{104}$Ru, $^{148}$Nd, $^{186}$Pt, $^{196}$Pt, $^{120}$Xe, $^{126}$Xe, $^{154}$Gd and $^{156}$Dy. The X(5) and X(3) model predictions are also shown for comparison.

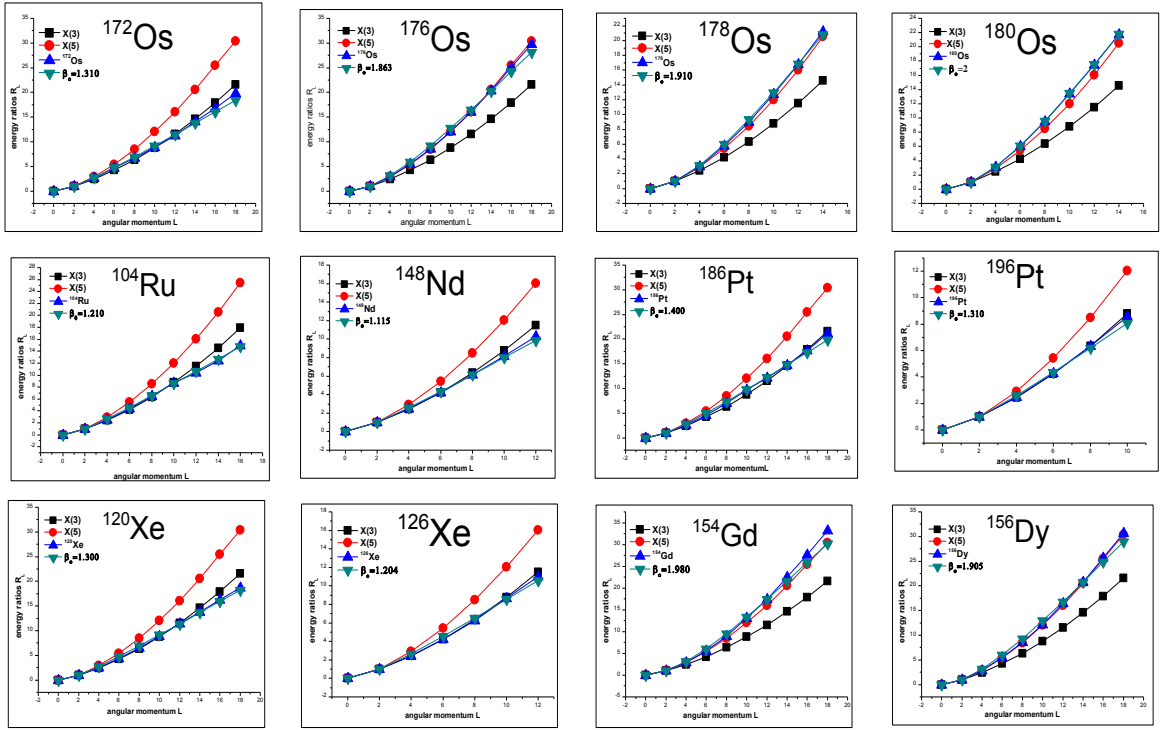

## V. Conclusion

In this study a $\gamma$-rigid solution of the Bohr Hamiltonian is obtained by solving the $\beta$ part with a Davidson potential. This model is called X(3)-D. The energy eigenvalues and wave functions are obtained by using an analytic method which has been developed by Nikiforov and Uvarov. B(E2) transition rates are also calculated. The X(3)-D model spans the path from the U(5) symmetry to the SU(3) symmetry involving a free parameter, $\beta_o$. Applying a variational procedure we point out that it is possible to suggest the X(3) model as a possible candidate for a critical point symmetry. The X(3)-D solution gives us the opportunity to search the experimental data on the path from the U(5) symmetry to the SU(3) symmetry, finding examples of nuclei corresponding to various values of $\beta_o$.

**Acknowledgement:** The authors acknowledge financial support by the Scientific and Technical Research Council of Turkey (TUBITAK), under the project number TBAG (112T754).

**Table I.** Energy levels of the X(3)-D model corresponding to different $\beta_o$ values. Ground state band, $\beta_1$ and $\beta_2$ band energies are normalized to the $2_{1,0}$ band lowest excited state. The X(3) model predictions and variational procedure for energy levels are also placed for comparison reason. See Section IV for further discussion.

| $\beta_o$ | 0 | 1 | 1.5 | 2 | ∞ | $\beta_{o,m}$ | var | X(3) |
|---|---|---|---|---|---|---|---|---|
| $L_{s,n}$ | | | | | | | | |
| $0_{1,0}$ | 0.00 | 0.00 | 0.00 | 0.00 | 0.00 | | | 0.00 |
| $2_{1,0}$ | 1.00 | 1.00 | 1.00 | 1.00 | 1.00 | | | 1.00 |
| $4_{1,0}$ | 2.13 | 2.48 | 2.90 | 3.14 | 3.33 | 1.10 | 2.57 | 2.44 |
| $6_{1,0}$ | 3.27 | 4.07 | 5.23 | 6.10 | 7.00 | 1.24 | 4.62 | 4.23 |
| $8_{1,0}$ | 4.42 | 5.71 | 7.79 | 9.60 | 12.00 | 1.35 | 7.13 | 6.35 |
| $10_{1,0}$ | 5.58 | 7.36 | 10.45 | 13.46 | 18.33 | 1.44 | 10.05 | 8.78 |
| $12_{1,0}$ | 6.73 | 9.02 | 13.19 | 17.56 | 26.00 | 1.53 | 13.47 | 11.52 |
| $14_{1,0}$ | 7.88 | 10.69 | 15.96 | 21.82 | 35.00 | 1.61 | 17.30 | 14.57 |
| $16_{1,0}$ | 9.04 | 12.37 | 18.76 | 26.19 | 45.33 | 1.68 | 21.59 | 17.91 |
| $18_{1,0}$ | 10.19 | 14.05 | 21.58 | 30.65 | 57.00 | 1.76 | 26.30 | 21.56 |
| $\beta_o$ | 0 | 1 | 1.5 | 2 | | $\beta_{o,m}$ | var | X(3) |
| $L_{s,n}$ | | | | | | | | |
| $0_{2,1}$ | 2.00 | 2.92 | 5.01 | 8.30 | | | | 2.87 |
| $2_{2,1}$ | 3.00 | 3.92 | 6.01 | 9.30 | | | | 4.83 |
| $4_{2,1}$ | 4.13 | 5.40 | 7.90 | 11.44 | | | | 7.37 |
| $6_{2,1}$ | 5.27 | 6.99 | 10.24 | 14.40 | | | | 10.29 |
| $8_{2,1}$ | 6.42 | 8.63 | 12.80 | 17.91 | | | | 13.57 |
| $10_{2,1}$ | 7.58 | 10.28 | 15.46 | 21.77 | | | | 17.18 |
| $12_{2,1}$ | 8.73 | 11.94 | 18.20 | 25.87 | | | | 21.14 |
| $\beta_o$ | 0 | 1 | 1.5 | 2 | | $\beta_{o,m}$ | var | X(3) |
| $L_{s,n}$ | | | | | | | | |
| $0_{3,2}$ | 4.00 | 5.84 | 10.02 | 16.61 | | | | 7.65 |
| $2_{3,2}$ | 5.00 | 6.84 | 11.02 | 17.61 | | | | 10.56 |
| $4_{3,2}$ | 6.13 | 8.32 | 12.91 | 19.75 | | | | 14.19 |
| $6_{3,2}$ | 7.27 | 9.91 | 15.25 | 22.70 | | | | 18.22 |
| $8_{3,2}$ | 8.42 | 11.55 | 17.81 | 26.21 | | | | 22.62 |

**Table II.** $B(E2; L_f \to L_i)$ rates for the X(3)-D model corresponding to different $\beta_o$ values for the ground state band, $\beta_1$ and $\beta_2$ bands. The B(E2) rates are normalized to the $B(E2; 2_{1,0} \to 0_{1,0})$ transition rate value from the lowest excited state to the ground state. The X(3) model predictions are placed for comparison reason.

| $\beta_o$ | 0 | 1 | 1.5 | 2 | 5 | X(3) |
|---|---|---|---|---|---|---|
| $L_i \to L_s$ | | | | | | |
| $2_{1,0} \to 0_{1,0}$ | 1.000 | 1.000 | 1.000 | 1.000 | 1.000 | 1.000 |
| $4_{1,0} \to 2_{1,0}$ | 2.377 | 1.912 | 1.638 | 1.522 | 1.432 | 1.889 |
| $6_{1,0} \to 4_{1,0}$ | 3.803 | 2.864 | 2.180 | 1.854 | 1.584 | 2.489 |
| $8_{1,0} \to 6_{1,0}$ | 5.237 | 3.854 | 2.759 | 2.185 | 1.669 | 2.914 |
| $10_{1,0} \to 8_{1,0}$ | 6.677 | 4.862 | 3.370 | 2.538 | 1.729 | 3.238 |
| $12_{1,0} \to 10_{1,0}$ | 8.117 | 5.878 | 3.998 | 2.910 | 1.777 | 3.495 |
| $14_{1,0} \to 12_{1,0}$ | 9.557 | 6.900 | 4.638 | 3.298 | 1.821 | 3.707 |
| $16_{1,0} \to 14_{1,0}$ | 10.997 | 7.926 | 5.287 | 3.698 | 1.862 | 3.885 |
| $18_{1,0} \to 16_{1,0}$ | 12.440 | 8.952 | 5.941 | 4.105 | 1.903 | 4.037 |
| | | | | | | |
| $2_{2,1} \to 0_{2,1}$ | 1.667 | 1.613 | 1.452 | 1.306 | 1.058 | 0.806 |
| $4_{2,1} \to 2_{2,1}$ | 3.203 | 2.585 | 2.179 | 1.926 | 1.515 | 1.401 |
| $6_{2,1} \to 4_{2,1}$ | 4.693 | 3.544 | 2.700 | 2.256 | 1.676 | 1.824 |
| $8_{2,1} \to 6_{2,1}$ | 6.163 | 4.542 | 3.261 | 2.571 | 1.764 | 2.155 |
| $10_{2,1} \to 8_{2,1}$ | 7.623 | 5.556 | 3.860 | 2.907 | 1.826 | 2.424 |
| $12_{2,1} \to 10_{2,1}$ | 9.077 | 6.580 | 4.481 | 3.267 | 1.875 | 2.651 |
| $14_{2,1} \to 12_{2,1}$ | 10.530 | 7.606 | 5.120 | 3.646 | 1.918 | |
| $16_{2,1} \to 14_{2,1}$ | 11.980 | 8.635 | 5.766 | 4.038 | 1.959 | |
| $18_{2,1} \to 16_{2,1}$ | 13.427 | 9.663 | 6.418 | 4.440 | 2.000 | |
| $20_{2,1} \to 18_{2,1}$ | 14.877 | 10.695 | 7.075 | 4.850 | 2.041 | |
| $22_{2,1} \to 20_{2,1}$ | 16.323 | 11.726 | 7.733 | 5.266 | 2.083 | |
| | | | | | | |
| $2_{3,2} \to 0_{3,2}$ | 2.333 | 2.229 | 1.908 | 1.616 | 1.117 | 0.735 |
| $4_{3,2} \to 2_{3,2}$ | 4.027 | 3.260 | 2.722 | 2.333 | 1.599 | 1.205 |
| $6_{3,2} \to 4_{3,2}$ | 5.583 | 4.224 | 3.220 | 2.661 | 1.768 | 1.542 |
| $8_{3,2} \to 6_{3,2}$ | 7.090 | 5.232 | 3,762 | 2,957 | 1,860 | 1.812 |

| | | | | | |
|---|---|---|---|---|---|
| $10_{3,2} \to 8_{3,2}$ | 8.573 | 6.253 | 4.351 | 3.276 | 1.923 |
| $12_{3,2} \to 10_{3,2}$ | 10.043 | 7.282 | 4.966 | 3.623 | 1.973 |
| $14_{3,2} \to 12_{3,2}$ | 11.503 | 8.313 | 5.599 | 3.992 | 2.016 |
| $16_{3,2} \to 14_{3,2}$ | 12.963 | 9.346 | 6.244 | 4.378 | 2.057 |
| $18_{3,2} \to 16_{3,2}$ | 14.417 | 10.377 | 6.894 | 4.775 | 2.097 |
| $20_{3,2} \to 18_{3,2}$ | 15.870 | 11.411 | 7.551 | 5.182 | 2.137 |
| $22_{3,2} \to 20_{3,2}$ | 17.320 | 12.446 | 8.211 | 5.594 | 2.178 |